\documentclass{acm_proc_article-sp}
\usepackage{amsmath}
\usepackage{multirow}
\usepackage{paralist}
\usepackage{graphicx}
\usepackage{enumitem}
\usepackage{mymacros}
\usepackage{natbib}

\setcitestyle{round,aysep={},yysep={;}}
\begin{document}
\bibliographystyle{agsm_nourl}

\title{Flexible Online Repeated Measures Experiment}

\numberofauthors{4} 
\author{
\alignauthor
Yu Guo\\
       \affaddr{Microsoft}\\
       \affaddr{One Microsoft Way}\\
       \affaddr{Redmond, WA 98052}\\
       \email{yguo@microsoft.com}
\alignauthor
Alex Deng\\
       \affaddr{Microsoft}\\
       \affaddr{One Microsoft Way}\\
       \affaddr{Redmond, WA 98052}\\
       \email{alexdeng@microsoft.com}
}

\date{20 Aug 2014}

\maketitle
\begin{abstract}
Online controlled experiments, now commonly known as A/B testing, are crucial to causal inference and data driven decision making in many internet based businesses. While a simple comparison between a treatment (the feature under test) and a control (often the current standard), provides a starting point to identify the cause of change in Key Performance Indicator (KPI), it is often insufficient, as the change we wish to detect may be small, and inherent variation contained in data may obscure movements in KPI. To have sufficient power to detect statistically significant changes in KPI, an experiment needs to engage a sufficiently large proportion of traffic to the site, and also last for a sufficiently long duration. This limits the number of candidate variations to be evaluated, and the speed new feature iterations. We introduce more sophisticated experimental designs, specifically the repeated measures design, including the crossover design and related variants, to increase KPI sensitivity with the same traffic size and duration of experiment. In this paper we present  FORME (Flexible Online Repeated Measures Experiment), a flexible and scalable framework for these designs. We evaluate the theoretic basis, design considerations, practical guidelines and big data implementation. We compare FORME to an existing methodology called mixed effect model and demonstrate why FORME is more flexible and scalable. We present empirical results based on both simulation and real data. Our method is widely applicable to online experimentation to improve sensitivity in detecting movements in KPI, and increase experimentation capability.
\end{abstract}



\vfill\eject 

\section{Introduction}

Many recent publications attest to the power of using online A/B testing as the golden rule for making causal inference in web facing companies large and small. By random assignment of feature to otherwise balanced groups of users and measuring subsequent changes in user behavior, A/B testing isolates effect of feature change, i.e. the treatment effect from extraneous sources of variance. 

To perform statistical inference in both point estimation and hypothesis testing for the treatment effect, while controlling type I error at pre-specified level, we would desire lower type II error, or equivalently, higher powered experiments. That is, we wish to be able to detect the effect when there is any. Running under powered experiments have many perils. Not only would we miss potentially beneficial effects, we may also get false confidence about lack of negative effects. Statistical power increases with larger effect size, and smaller variances. Let us look at these aspects in turn. 

While the actual effect size from a potential new feature may not be known, we generally select a size that makes business sense, i.e. one that justifies the cost of feature development and ongoing maintenance of the code base. Dramatic features that drastically alter user behavior and get reflected in KPI as large effect sizes are few and far in between. Often the candidate feature has but a small effect on the KPI. Nonetheless, by accumulating a portfolio of small changes, a business can achieve big business success. Quote from Rule \#2 of \citet{Kohavi2014SevenRules}, winning is done inch by inch. This is especially true for mature web facing businesses where most low hanging fruits were picked already.

In general one expects variance to decrease with increased sample size. But this is not always true. For online business, at first glance it may seem that the number of visitors may be large, and with a casual look people may think the power to detect any change is large. In reality, however, intrinsic variation between users is large and may obscure the small movement in KPI. Variation in measured treatment effect comes from various sources. Exogenous to the treatment itself includes user to user variation, e.g. some users from slower internet connection would always have slower page load time regardless of what experiments are run. Variance for some metrics does not decrease over time, instead they plateau after some period of time (say, two weeks), and running longer experiments no longer results in corresponding benefits \citep[see][Section 3.4]{puzzlingOutcomes}. This poses a limitation to any online experimentation platform, where fast iterations and testing many ideas can reap the most rewards.

\subsection{Motivation} 
\label{sub:motivation}

To improve sensitivity of measurement, apart from accurate implementation and increase sample size and duration, we can employ statistical methods to reduce variance. Using the user's pre-experiment behavior as a baseline for his/her post-experiment behavior, we can reduce the variance in measured treatment effect. The experiment setup in a two-week experiment is shown in Table \ref{diagram_3_designs}. The typical A/B test is illustrated in the first row. In the past we have used regression to reduce variance (CUPED: Controlled Experiments Using Pre-Experiment Data, see \citet{deng2013cuped}) and have achieved good results, e.g. reducing variance in number of queries per unique user in a given time period by 40-50\%. CUPED has the benefit of having readily available baseline data ``for free''. This improvement is performed with existing design, using the ``free'' data as covariates only in the analysis stage. CUPED is in fact a form of repeated measures design, where multiple measures on the same subjects are taken over time. In particular, in the pre-experiment stage, all users received the default feature C (control) and none received the new feature T (treatment). 
\begin{table}[h]
\centering
\resizebox{\columnwidth}{!}{%
\begin{tabular}{rclcc}

\multicolumn{1}{l}{}                    & \multicolumn{1}{l}{} & \multicolumn{1}{c}{Pre-}       & \multicolumn{2}{c}{\multirow{2}{*}{Experiment}} \\
\multicolumn{1}{l}{}                    & \multicolumn{1}{l}{} & \multicolumn{1}{c}{Experiment} & \multicolumn{2}{c}{}                            \\ \cline{3-5} 
\multicolumn{1}{l}{}                    & Groups               & \multicolumn{1}{c}{Week0}      & Week1                           & Week2         \\ \hline
\multirow{2}{*}{\textbf{A/B Test}}       & 1                    &                                & \multicolumn{2}{c}{- T -}                       \\
                                        & 2                    &                                & \multicolumn{2}{c}{- C -}                       \\ \hline
\multirow{2}{*}{\textbf{CUPED}}         & 1                    & \multicolumn{1}{c}{C}          & \multicolumn{2}{c}{- T -}                       \\
                                        & 2                    & \multicolumn{1}{c}{C}          & \multicolumn{2}{c}{- C -}                       \\ \hline
\multirow{2}{*}{\textbf{Parallel}}      & 1                    &                                & \multicolumn{1}{c|}{T}          & T             \\
                                        & 2                    &                                & \multicolumn{1}{c|}{C}          & C             \\ \hline
\multirow{2}{*}{\textbf{Crossover}}     & 1                    & \multicolumn{1}{c}{}           & \multicolumn{1}{c|}{T}          & C             \\
                                        & 2                    & \multicolumn{1}{c}{}           & \multicolumn{1}{c|}{C}          & T             \\ \hline
\multirow{4}{*}{\textbf{Re-Randomized}} & 1                    &                                & \multicolumn{1}{c|}{C}          & C             \\
                                        & 2                    &                                & \multicolumn{1}{c|}{C}          & T             \\
                                        & 3                    &                                & \multicolumn{1}{c|}{T}          & C             \\
                                        & 4                    &                                & \multicolumn{1}{c|}{T}          & T            
\end{tabular}}
\caption{Repeated Measures Designs}
\label{diagram_3_designs}
\end{table}\\
In this paper we extend the idea further by employing the repeated measures design in different stages of treatment assignment. The traditional A/B test can be analyzed using the repeated measures analysis, reporting a ``per week'' treatment effect, as show in row 3 ``parallel'' design in table \ref{diagram_3_designs}. The two week experiment can be considered to be conducted in two periods, even though users received the same treatment assignment during both periods. In one of the new designs, the ``crossover'' design, in contrast, we swap treatment assignment half way through the experiment (row 4 in table \ref{diagram_3_designs}). Each user will be exposed to both versions of the treatments, instead of only one of the two in the usual A/B testing scenario. In sequence, a user will receive either T followed by C, or C followed by T, with the flight re-assignment happening at the same moment for all users. Instead of randomizing treatments to users, we random treatment sequences (TC or CT) to users. This way each user serves as his/her own control in the measurement. In fact, the crossover design is a type of repeated measures design commonly used in biomedical research to control for within-subject variation. We also discuss practical considerations to repeated measures design, with variants to the crossover design to study the carry over effect, including the ``re-randomized'' design (row 5 in table \ref{diagram_3_designs}). 

\subsection{Main Contributions} 
\label{sub:main_contributions}
In this paper, we propose a framework called FORME (Flexible Online Repeated Measures Experiment). We made contributions in both novel application and new methodology.
\begin{compactitem}
\item Novel applications. We propose different experiment designs with repeated measurement. We demonstrate through real examples the value of these new designs comparing to traditional A/B test. Methods for model assumption checking is also presented. We also compare different designs for practical use and propose a general workflow for practitioners.
\item New Methodology. We review standard repeated measures models in the framework of mixed effect models. We present a new method to fit the model that is scalable to big data. Our method is flexible in the sense that it makes far less assumptions than traditional method based on mixed effect model \citep{Bates2012}. It naturally handles missing data without missing at random assumption (common in online experimentation) and still provides unbiased average treatment effect estimation when mixed effect model fails. FORME can fit different types of repeated measures models under the same framework. It also can be applied to metrics beyond those defined as a simple average, such as metrics defined as a function of other metrics.  
\end{compactitem}


\section{Illustration of FORME} 
\label{sec:Background}



In this sections we will take a close look at several designs, with a treatment and a control, and with experiments carried out over several periods. Many common online metrics display different patterns between weekdays and weekends. Therefore experiments at Bing and many large IT companies, in general are run for at least a full week to account for the difference between weekdays and weekends. In the following section we assume the minimum experimentation ``period'' to be one full week, and may extend to up to two weeks. To facilitate our illustration, in all the derivation in this section we assume all users appear in all periods, i.e. no missing measurement. We also restrict ourselves to metrics that are defined as simple average and assume treatment and control have the same sample size. We further assume treatment effects for each subjects are fixed. We emphasis this is just for illustration purpose and our method does not rely on these assumptions and we describe how we handle missing data and more complicated metrics in Section~\ref{sec:missing_values}. Impatient reader who are familiar with repeated measures analysis might jump over to Section~\ref{sec:theory} to see details of FORME's model assumptions and comparison to linear mixed effect model. 

Denote the metric value mean in the treatment group as $\mu_T$, and that in control as $\mu_C$. We are interested in the average treatment effect (ATE) $\delta = \mu_T-\mu_C$ which is a fixed effects in the model in this section.  This way, various designs considered can be examined in the same framework and easily compared. 

We will proceed to show, with theoretical derivations, that given the same total traffic
\begin{compactitem}
\item Variance using CUPED $\le$ T-Test
\item With CUPED: Variance in parallel design $\le$ Cumulative Design 
\item Variance in Crossover design $\le$ Parallel Design
\end{compactitem}

Denote observed sample values in the treatment groups and time periods as $\vec{X}$, and their means $\vec{\beta}$. Note that $\vec{X}$ is a vector of metric values $\xbar_i$ for different time periods indexed by $i$, and the treatment effect $\delta$ can be formulated as a function of $\vec{\beta}$ depending on model specification. Under the central limit theorem (CLT), with sufficiently large samples $\vec{X}$ is asymptotically normal
\begin{equation*}
       \vec{X} \sim N(\vec{\beta}, \Sigma).
\end{equation*}
The likelihood of $\vec{\beta}$ given observed data is then
\begin{equation*}
       L = \frac{1}{\sqrt{2\pi} |\Sigma|^{\frac{1}{2}}}\exp{\big(\frac{1}{2}(\vec{X}-\vec{\beta})^T\Sigma^{-1}(\vec{X}-\vec{\beta})\big)}
\end{equation*}
To get maximum likelihood estimates (MLE) of $\vec{\beta}$, denoted by $\hat{\beta}$, we seek to minimize -2$\log(\text{Likelihood})$
\begin{equation*}
       l = -\frac{1}{2}(\vec{X}-\vec{\beta})^T\Sigma^{-1}(\vec{X}-\vec{\beta}) + const
\end{equation*}
Solving ${\frac{\partial l}{\partial \vec{\beta}} }=0$ gives MLE of $\vec{\beta}$. And its variance-covariance matrix is
\begin{equation*}
Var(\hat{\beta})=\left[{\frac{\partial^2 l}{\partial \vec{\beta} \partial \vec{\beta}^T} }\right]^{-1} = 1/I(\vec{\beta}) 
\end{equation*}
where Fisher Information 
\begin{equation*} 
I(\vec{\beta})=  - E\Big[\big(\frac{\partial}{\partial \vec{\beta}} log f(X|\vec{\beta}) \big)^T log f(X|\vec{\beta}) \Big| \vec{\beta} \Big].
\end{equation*}
In the following sections we will explicitly model the mean $\vec{\beta}$ as a function of other parameters $\vec{\beta}(\vec{\lambda})$, one of the components is treatment effect $\delta$, and study expected variance of the MLEs of $\vec{\lambda}$. In fact this is simply:
\begin{align*}
Var(\hat{\lambda}) & = I(\vec{\lambda})^{-1} \\
&= \big [ \big( \frac{\partial \beta}{\partial \lambda}\big)^T \Sigma ^{-1} E \big [ (X-\beta) (X-\beta) ^ T \big] \Sigma ^{-1} \frac{\partial \beta}{\partial \lambda} \big ] ^{-1}\\
& =  \big [ \big( \frac{\partial \beta}{\partial \lambda}\big)^T \Sigma ^{-1}  \frac{\partial \beta}{\partial \lambda} \big ] ^{-1}
\end{align*} 
Coefficient of variation (CV) defined as the mean over standard deviation of a metric, determines the sensitivity or the power of the experiment, given the same sample size. To study sensitivity or power of various experimental designs, once we have established that effect size remains relatively stable across different measuring periods, we can then focus on variation of estimated effect size solely. Specifically, the diagonal cell in $Var(\hat{\lambda})$ corresponding to treatment effect $\hat{\delta}$ gives its variance, and is our main focus in the following of this section. 

Analysis from randomized two-group experiments employs the two sample t-test under the usual A/B testing scenario. As a gentle introduction we will first look at the t-test using this notation. 

\subsection{Two Sample T-test} 
\label{sub:two_sample_t_test}

Let $\xbar$ denote the observed average metric value in control group and $\ybar$ denote that in the treatment group. Since users are randomly assigned into either treatment or control group, $\xbar$ and $\ybar$ are thus independent. For simplicity of notation, we assumed variance in the two group to be equal. Given large enough sample size, under CLT, and plug in observed sample variances, we have:
\begin{equation*}
      \left[ \begin{smallmatrix}{}C\\-\\T \end{smallmatrix} \right] : \left[\begin{smallmatrix}{}\xbar\\\ybar\end{smallmatrix}\right]
       \sim N \Big(
       \left[\begin{smallmatrix}{}\mu\\\delta + \mu\end{smallmatrix}\right] , \left[\begin{smallmatrix}{}s_{X}^{2} & 0\\0 & s_{Y}^{2}\end{smallmatrix}\right]
         \Big)
\end{equation*}
where $\mu$ is mean metric value in the control group, $\delta$ is the treatment effect compared to the control group, and $s_{X}^2$ and $s_Y^2$ are variances of $\xbar$ and $\ybar$ respectively. Here $\vec{\lambda}= (\mu, \delta)$, and the $-2log\text{Likelihood}$, denoted by $l$, of parameter vector $\vec{\beta}(\vec{\lambda})=(\mu, \delta)^T$ given observed data is then 
\begin{equation*}
       l =  \left[\begin{smallmatrix}{}X - \mu\\Y - \delta - \mu\end{smallmatrix}\right] ^T \left[\begin{smallmatrix}{}s_{X}^{2} & 0\\0 & s_{Y}^{2}\end{smallmatrix}\right] ^{-1} \left[\begin{smallmatrix}{}X - \mu\\Y - \delta - \mu\end{smallmatrix}\right] + const
\end{equation*}
Solving for MLE of $\delta$ and obtain its variance as 
\begin{equation}\label{ttest}
      Var(\hat{\delta})|_{TTest} =s_X^2+s_Y^2
\end{equation}
It is simply the sum of variances from the treatment and control groups, which is the asymptotic variance of $\xbar-\ybar$.

\subsection{Use Pre-experiment Data for Variance Reduction} 
\label{sub:use_pre_experiment_value_for_variance_reduction}
At the analysis level, different models seek to explain the amount of variation in observed data, which may come from intrinsic, within-user difference, as well as variation introduced by differential treatment. For example, users that connect through broadband tend to have faster page load time than people using dial-up connection. This difference exists regardless of which treatment conditions the users are exposed to, and is thus irrelevant when measuring difference introduced by different treatments. As a result, the measurements on the same users over time tend be positively correlated. 

CUPED  and previous work has established that by including covariates that are unrelated to the treatment, we can improve sensitivity and reduce variance of estimated treatment effect. Specifically, the users' pre-experiment behaviors servers as a good baseline for their behavior during the experiment. By including pre-experiment data as a covariate in the regression model for treatment effect, we can reduce the variance of the estimated treatment effect. 

Denote the pre-experiment average metric value to be $\xbar_0$ and $\ybar_0$ for the later control and treatment groups respectively. By CLT
\begin{equation*}
\left[\begin{smallmatrix}{}C\\C\\-\\C\\T\end{smallmatrix}\right]: 
\left[\begin{smallmatrix}{}\xbar_{0}\\ \xbar_{1}\\ \ybar_{0}\\ \ybar_{1}\end{smallmatrix}\right]
\sim N \Big(
\left[\begin{smallmatrix}{}\mu\\\mu + \theta\\\mu\\\delta + \mu + \theta\end{smallmatrix}\right] ,  \left[\begin{smallmatrix}{}\Sigma & 0 \\0 & \Sigma \end{smallmatrix}\right]
  \Big), 
  \Sigma = \left[\begin{smallmatrix}{}s_{0}^{2} & \rho s_{0} s_{1}\\\rho s_{0} s_{1} & s_{1}^{2}\end{smallmatrix}\right]
\end{equation*}
where $\theta$ is the difference between the pre-experiment and experiment periods, i.e. the longitudinal effect, and $\rho$ is the correlation between the two periods. Here $\vec{\lambda} = (\mu, \delta, \theta)$. We assume correlation $\rho$ in both treatment and control groups to be the same for simplicity. Results do not dependent on this assumption. Even though the two treatment groups are still independent, metric value measured on the same group of users across different time periods are in general correlated. As we will later see, this correlation effectively reduces variances on $\hat{\delta}$. Similarly we can solve for MLEs from solving partial derivative of $l$ = 0 and derive variances for these estimates. 
\begin{equation}\label{2week_aa_ab}
       Var(\hat{\delta})|_{CUPED}= 2 s_{1}^{2} \left(1- \rho^{2} \right)
\end{equation}
It's easy to see \eqref{2week_aa_ab} has smaller variance of $\hat{\delta}$ than \eqref{ttest} by amount of $ 2 \rho^{2} s_{1}^{2}$. As users' behavior is usually consistent across time, i.e. with non-zero correlation $\rho$ among different time periods, this amount is positive. The amount of variance reduced is $ \rho^{2} $ that of the original variance. 

\subsection{Cumulative vs. Parallel Design} 
\label{sub:cumulative_vs_parallel_design}

Note that in the previous design we make no assumption on the duration of the pre-experiment and experiment periods. Empirical studies in \cite{deng2013cuped} have shown that using one-week pre-experiment data provides similar amount of variance reduction as using even longer durations. For simplicity, in practice we recommend using one-week such data. 

And we have mentioned that to capture the difference between weekday and weekends, we recommend running experiments for whole weeks, typically 14 days. Assuming treatment effect is the same across time, this gives us two ways of reporting treatment effects, i.e. reporting cumulative effects for the whole 14 days, and reporting weekly treatment effect as a weighted average between observed values in the two weeks. For the latter, using the same notation as above, we have
\begin{equation*}
\left[\begin{smallmatrix}{}C\\C\\-\\T\\T\end{smallmatrix}\right]: 
\left[\begin{smallmatrix}{}\xbar_{1}\\ \xbar_{2}\\ \ybar_{1}\\ \ybar_{2}\end{smallmatrix}\right]
\sim N \Big(
\left[\begin{smallmatrix}{}\mu\\\mu + \theta\\\delta + \mu\\\delta + \mu + \theta\end{smallmatrix}\right] , \left[\begin{smallmatrix}{}\Sigma & 0 \\0 & \Sigma \end{smallmatrix}\right]
  \Big), 
  \Sigma = \left[\begin{smallmatrix}{}s_{1}^{2} & \rho s_{1} s_{2}\\\rho s_{1} s_{2} & s_{2}^{2}\end{smallmatrix}\right]
\end{equation*}

We can solve for MLE and their variances. 
\begin{equation}\label{2week_ab_ab}
Var(\hat{\delta})|_{Parallel}= 2 \frac{s_{1}^{2} s_{2}^{2} \left(1 -  \rho^{2}\right)}{s_{1}^{2} + s_{2}^{2}- 2 \rho s_{1} s_{2}}
\end{equation}
For the former, if the metric value is strictly additive across time, an example being revenue, under our toy model where all users appear in both periods, the cumulative treatment effect would be $\tilde{\delta} = 2\delta$, since
\begin{align*}
  \left[ \begin{smallmatrix}{}C\\-\\T \end{smallmatrix} \right] :      \left[\begin{smallmatrix}{}\xbar_{1}+\xbar_{2}\\ \ybar_{1}+\ybar_{2}\end{smallmatrix}\right]
       &\sim N \Big(
       \left[\begin{smallmatrix}{}2\mu+\theta\\2\delta + 2\mu+\theta\end{smallmatrix}\right] ,\left[\begin{smallmatrix}{}\Sigma & 0 \\0 & \Sigma \end{smallmatrix}\right]
  \Big), \\ \Sigma&=Var(\xbar_1+\xbar_2). 
\end{align*}
Using \eqref{ttest}, variance for the MLE is
\begin{equation}\label{cumulative_ab}
       Var(\hat{\tilde{\delta}})|_{Cumulative} = 2Var(\xbar_1+\xbar_2) = 2(s_1^2+s_2^2+2\rho s_1 s_2)      
\end{equation}
Comparing coefficient of variation (CV) in \eqref{2week_ab_ab} to \eqref{cumulative_ab},
\begin{align*}
 & \frac{Var(\hat{\tilde{\delta}})}{4\delta^2} - \frac{Var(\hat{\delta})}{\delta^2}  
       = \frac{ (s_{1}+ s_{2})^{2} (s_{1}- s_{2})^{2}}{2\delta^2 (s_{1}^{2} + s_{2}^{2}- 2 \rho s_{1} s_{2})} 
       \ge 0
\end{align*}
Equality holds when the two periods have identical variance, i.e. $s_1 = s_2$. In other words, for additive metrics which variation over time is large, reporting weekly metrics alone will improve metric sensitivity. 

For non-additive metrics, such as ratio metrics like Click Through Rate (CTR), the derivation becomes more involved. In practice, also there is a lot of non-recurring users. We opted to show empirical results instead in results section. 

Careful readers may have noticed, this method makes a key assumption that treatment effect $\delta$ remains the same in the two weeks. To check this assumption, we can explicitly test for $\delta_1 = \delta_2$ by fitting the model this way:
\begin{equation*}
E\left[\begin{smallmatrix}{}\xbar_{1}\\ \xbar_{2}\\ \ybar_{1}\\ \ybar_{2}\end{smallmatrix}\right]
=
\left[\begin{smallmatrix}{}\mu\\\mu + \theta\\\delta_1 + \mu\\\delta_2 + \mu + \theta\end{smallmatrix}\right]
\end{equation*}
and test for the equivalence of MLEs $H_0: \hat{\delta_1} = \hat{\delta_2}$. The parallel design is appropriate if we fail to reject $H_0$.

\subsection{Crossover Design}
\label{sub:crossover_design}
Now with the preliminary background information setup, we then look at variation reduction achieved through the crossover design. 

The crossover design employs a similar idea to CUPED. Instead of using pre-experiment data as the baseline, in crossover experiments, each user is exposed to both treatments sequentially, while the order of treatment groups is determined randomly. Each user's behavior while he or she is on the control condition serves as a baseline for his or her behavior on the treatment condition. By accounting for within-user variation, analysis based on the crossover design also reduces variance for the estimated treatment effect. 

In causal inference, we often seek to eliminate any confounding factors and isolate the root cause of observed difference. Due to not observing the counterfactual in the potential outcome framework\citep{Rosenbaum1983,counterfac}, randomization is used to make the control group as the surrogate for counterfactual. This surrogate only works \emph{on average}. In reality often some imbalance in some observed or unobserved factors will remain. Crossover design uses each test subject as his or her own control, thus reducing the influence of confounding covariates, and achieve better sensitivity in estimating treatment effect. 

Distribution of observed sample averages is:
\begin{equation*}
\left[\begin{smallmatrix}{}C\\T\\-\\T\\C\end{smallmatrix}\right]: \left[\begin{smallmatrix}{}\xbar_{1}\\ \xbar_{2}\\ \ybar_{1}\\ \ybar_{2}\end{smallmatrix}\right]
\sim N \Big(
\left[\begin{smallmatrix}{}\mu\\\delta + \mu + \theta\\\delta + \mu\\\mu + \theta\end{smallmatrix}\right] , 
\left[\begin{smallmatrix}{}\Sigma & 0 \\0 & \Sigma \end{smallmatrix}\right]
  \Big), 
  \Sigma = \left[\begin{smallmatrix}{}s_{1}^{2} & \rho s_{1} s_{2}\\\rho s_{1} s_{2} & s_{2}^{2}\end{smallmatrix}\right]
\end{equation*}
Similarly, treatment effect estimate has variance
\begin{equation} \label{2week_ab_ba_crossover}
Var(\hat{\delta})|_{Crossover}= 2 \frac{s_{1}^{2} s_{2}^{2} \left(1-\rho^{2}\right)}{s_{1}^{2} + s_{2}^{2}+2 \rho s_{1} s_{2} }
\end{equation}
Comparing \eqref{2week_ab_ab} to \eqref{2week_ab_ba_crossover}, it is obvious that in the crossover design, treatment effect has smaller variance as long as the correlation $\rho$ is positive. Similar to CUPED, the amount of sensitivity improvement is determined by the size of $\rho$. The larger the correlation between time periods, the more improvement the crossover design has over the parallel design. The equivalence of treatment effect can be similarly checked as in section \ref{sub:cumulative_vs_parallel_design}.

\subsection{Absolute or Relative Change?}
\label{sub:absolute_or_relative_change}
So far in this paper we considered the absolute treatment difference $\delta = \mu_{T} - \mu_{C}$. In practice we measure thousands of metrics simultaneously. These metrics may have vastly different magnitude in their treatment effects. Even the same metric measured over different duration, or over different sample sizes may have different absolute $\delta$'s. This renders comparison of effect size across different experiments difficult. To overcome this difficulty, we often seek to measure percent delta, $\%\delta = \frac{\delta}{\mu_{C}} \cdot 100\%$. The relative change is less influenced by the base difference and is a more robust measure of treatment effect. In online experimentation we usually deal with hundreds of thousands of samples, therefore CLT still holds and relative change would still have asymptotic normality. The additive model described above can be readily adapted to model relative difference instead of absolute difference, by formulating the expected group means in the mixed variance-covariance structure model. For example, the crossover model with relative treatment effect can be written as:
\begin{equation*}
\left[\begin{smallmatrix}{}C\\T\\-\\T\\C\end{smallmatrix}\right]: \left[\begin{smallmatrix}{}\bar{X}_{1}\\\bar{X}_{2}\\\bar{Y}_{1}\\\bar{Y}_{2}\end{smallmatrix}\right]
\sim N \Big(
\left[\begin{smallmatrix}{}\mu\\\mu(1+\delta)+ \theta\\\mu(1 + \delta)\\\mu + \theta\end{smallmatrix}\right] , 
\left[\begin{smallmatrix}{}\Sigma & 0 \\0 & \Sigma \end{smallmatrix}\right]
  \Big), 
  \Sigma = \left[\begin{smallmatrix}{}s_{1}^{2} & \rho s_{1} s_{2}\\\rho s_{1} s_{2} & s_{2}^{2}\end{smallmatrix}\right]
\end{equation*}
Theoretic derivation to show variance reduction can be complex, but MLE estimates and their variances can be easily solved using numeric methods. 

\subsection{The Unified Theme}\label{sec:motif}
We illustrated different model designs of FORME. Careful readers might already noticed that the unified theme here is to study the joint distribution of $\vec{X}$ and $\vec{Y}$, which by central limit theorem is known to be multivariate normal. Each model specification maps to the mean vector $\vec{\beta}$ of this multivariate normal. Therefore for any mean vector based on a model specification, we can solve the MLE and estimate its variance using Fisher's Information. The difficulty, however, lies in how to estimate the covariance matrix in general case with presence of missing data and in general for metrics that are not defined as simple as average. For crossover design, in particular, we also need a way to decide whether we can safely assume the treatment effect in both periods are the same without any carry over effect. We will address these in details in Section~\ref{sec:carry_over_effect} and Section~\ref{sec:missing_values}. Section~\ref{sec:theory} explains why FORME is more flexible and scalable than existing method of fitting linear mixed effect model.

\section{Carry over effect} 
\label{sec:carry_over_effect}
The crossover design is not without concerns. An important assumption in crossover model is that the treatment effect remains the same in the experimental periods. Since test subjects randomly receive all combinations of treatments in sequence, different users will receive difference sequence or ``order'' the treatments. It is possible that the order in which users are exposed to treatments may change the effect. For an extreme example, suppose our treatment introduced a bug that results in severely negative user experience, and these group of users fail to revisit the website in the later crossover period, the treatment effect is then different in the two periods. We call this the carry over effect, as the users exposed to treatment first, and then the control later may behave differently from the other group. 

In some experiments where the treatment condition is less noticeable to the users, the expected treatment effect is small, and based on historical insight, it is safe to assume no carry over effect exists. Usually a ``wash-out'' period can be injected in between treatment periods. 
\subsection{Wash-out Period} 
\label{ssub:washout}

This approach calls for a ``wash-out'' period after the end of the first period, where all users will receive the control. Data from the wash-out period can be analyzed in similarly in linear mixed model to estimate the carry over effect and subsequently inform the design of later stage. We leave this as an exercise to the reader.

\subsection{Estimate Carry over Effect} 
\label{sub:estimate_carry over_effect}
In the crossover design where only two groups are allocated, it is not hard to see that potential carry over effect is confounded with the week to week difference of treatment effect. Using only the crossover model, we can only measure one of these two effects. As an alternative, at the cost of less efficiency gain over the traditional non-crossover design, we may estimate the carry over effect explicitly using the following 4-group design.

In a 4-group re-randomized design, we conduct the experiment over two periods, and split the users into four equally sized groups, one receiving controls in both periods, one receiving treatments in both, and one receiving treatment followed by control, and the last receiving control followed by treatment, we can then tease apart carry over effect and treatment effect. We call this the re-randomized design, as it is equivalent to having another round of user randomization between the first and the second period. Using notation from the linear mixed model, the model considering a potential carry over effect $\alpha$ is then 
\begin{align*}
\left[\begin{smallmatrix}{}C\\T\\-\\T\\C\\-\\C\\C\\-\\T\\T\end{smallmatrix}\right]: \left[\begin{smallmatrix}{} \xbar_{1}\\ \xbar_{2}\\ \ybar_{1}\\ \ybar_{2}\\ \zbar_{1}\\ \zbar_{2}\\ \wbar_{1}\\ \wbar_{2}\end{smallmatrix}\right]
&\sim N \Big(
\left[\begin{smallmatrix}{}\mu\\\delta + \mu + \theta\\\delta + \mu\\\alpha+\mu + \theta\\\mu\\\mu + \theta\\\delta + \mu\\\delta + \mu + \theta\end{smallmatrix}\right] , \left[\begin{smallmatrix}{}\Sigma & 0 & 0 & 0\\0 & \Sigma & 0 & 0\\0 & 0 & \Sigma & 0\\0 & 0 & 0 & \Sigma\end{smallmatrix}\right]
  \Big), \\
  \Sigma &= \left[\begin{smallmatrix}{}s_{1}^{2} & \rho s_{1} s_{2}\\\rho s_{1} s_{2} & s_{2}^{2}\end{smallmatrix}\right]
\end{align*}
The carry over effect is in the group that received first treatment and then reverted back to control in the second stage. Using observed data we can estimate carry over effect $\alpha$ as an additional term in $\vec{\beta}$. When $\alpha$ is not statistically significant under pre-specified type I error cutoff (usually 0.05), it is safe to drop the term $\alpha$ from $\vec{\beta}$, and re-fit the model. This way, we gain one more degree of freedom and thus reduces variances in the MLEs. 

This approach enables direct estimation of carry over effect. It can also be considered a hybrid approach between the crossover and parallel design, as half of the users received crossed treatments, and half received the same treatments in the two periods. It is not hard to see this design will achieve sensitivity improvement between the crossover and parallel designs. 


\section{Missing Values and Metrics Beyond Average}\label{sec:missing_values}
\subsection{Loss of follow-up and intent to treat} 
\label{sub:loss_of_follow-up_and_intent_to_treat}
Loss of follow-up is a common term in clinical studies. It refers to patients who were active participants during some period of the trial, but stopped participating at some point of follow-up. This can lead to incomplete study results and potential bias when the attrition is not random. Intention-to-treat analysis is commonly employed, when the subjects' initial assigned  treatment is used regardless of actual received treatment. In online A/B testing the idea is similar; users are assigned to treatment groups at some point in time before the experiment starts, often by user id, but may or may not appear in the actual duration of the experiment. This missing pattern is far from random, therefore methods that rely on strong MCAR assumption (missing completely at random) are not appropriate and even MAR (missing at random) assumption is questionable as it requires missing pattern to be random conditioned on observed covariates. One way to see measurements are not missing at random is to realize infrequent users are more likely to have missing values and the absence in a specific time window can still provide information on the user behavior and in reality there might be other factors causing user to be missing that are not even observed. Instead of throwing away data points where user appeared in only one period and is exposed to only one of the two treatments, in practice, we included an additional indicator for whether or not the user appeared in the study in the period. 

Specifically, we use an additional indicator for the presence/absence status of a user in each experimentation period. For user $j$ in period $i$, let $I_{ij} = 1$ if user $j$ appears in period $i$, and $0$ otherwise. For each user $j$ in period $i$, instead of one scalar metric value $(X_{ij})$, we augmented it into a vector $(I_{ij}, X_{ij})$. When $I_{ij}=0$, i.e. user is missing, we define $X_{ij}=0$. Under this simple augmentation, the metric value $\xbar_i$ for period $i$, taking average over those non-missing measurements, is the same as $\frac{\sum_k X_{ik}}{\sum_k I_{ik}}$. In this connection, to obtain MLE and its variance, we only need to estimate the covariance matrices for each group across time periods, i.e. 
\begin{align*}
\cov(\overline{X_i}, \overline{X_{i'}}) &= \cov\left(\frac{\sum_k X_{ik}}{\sum_k I_{ik}},\frac{\sum_{k'} X_{i'k'}}{\sum_{k'} I_{i'k'}}\right) \\ 
& = \cov\left( \frac{\overline{X_{i}}}{\overline{I_i}}, \frac{\overline{X_{i'}}}{\overline{I_{i'}}} \right)
\end{align*}
where the last equality is by dividing both numerator and denominator by the same total number of users who have \emph{ever} appeared in the experiments. Thanks to the central limit theorem, the vector $(\overline{I_i},\overline{X_i},\overline{I_{i'}},\overline{X_{i'}})$ is also asymptotically normal. Plugging in observed sample means and covariance matrix, $Cov(\overline{X_i}, \overline{X_{i'}})$ can be trivially computed with the $delta$-method; see \citet[Appendix B]{deng2013cuped} for a similar example; also see \citep{VanderVaart2000} for a text book treatment of the $delta$-method.

\subsection{Metrics Beyond Average} 
\label{sub:page_level_metrics}
Treatment groups are assigned to users, but not all metrics are simple averages across users. We can define a metric as a function of other metrics. One important family of metrics is page level metrics such as click through rate. Page level metrics use number of page-views as their denominator. At first glance it might look like just a simple average. Since treatments are assigned to users (the independent unit), page-views are therefore not independent. Considering it as simple average over page level measurements this needs extra care. A better approach is to see this as a ratio of two user level metrics: clicks per user and page-views per user. 
\begin{equation*}
 \frac{\sum_{user_i}Clicks_i}{\sum_{user_i}Pages_i} = \frac{\sum_{user_i}Clicks_i/\sum_{user_i}I_i}{\sum_{user_i}Pages_i/\sum_{user_i}I_i},
\end{equation*}
where $\sum_{user_i}I_i$ is the the count of appeared users. The same $delta$-method mentioned in Section~\ref{sub:loss_of_follow-up_and_intent_to_treat} naturally extends here, with slightly more complicated formula. Since $delta$-method applies in general to any continuous function, we can handle any metric that is defined as a continuous function of other metrics.

\section{Flexible and Scalable Repeated Measures Analysis via FORME} 
\label{sec:theory}
\subsection{Review of Existing Methods}
It is common to analyze data from repeated measures design with the repeated measures ANOVA model and the F-test, under certain assumptions, such as normality, sphericity (homogeneity of variances in differences between each pair of within-subject values), equal time points between subjects, and no missing data. Such assumptions in general do not hold for large-scale online experiments, where the assignment of users into different treatment group may not be completely balanced. 

A generally more applicable method is to analyze the data using the linear mixed effect model, for which complete balance is not necessary \citep{Bates2012}. In particular, a linear mixed effect model treat each measurement of a subject as a data point, and model the measurement as 
\begin{equation*}
Y = \theta + \alpha X+\beta Z+\epsilon
\end{equation*}
Here $\theta$ is the global mean and $\alpha$ stands for the vector of all deterministic fixed effects while $\beta$ is the vector of all random effects and $\epsilon$ is noise. $X$ and $Z$ are covariates in the model. In our cases they are indicators of treatment assignment, periods of the measurement, user id, and any other covariate. As an example, one possible model for repeated measures using lme4's formula syntax \citep{Bates2012,Bates2012a} is 
\begin{align*}
Y \sim 1 + IsTreatment + Period + (1|UserID),
\end{align*}
where the only difference of this model to the usual linear model behind two sample test is the extra random effect(clustered by UserID) to model user ``baseline''. More complicated models exist to further model interaction and joint random effects. 

Random effect makes modeling within-subject variability possible. In repeated measures data, users might appear in multiple periods, represented as multiple rows in the dataset. As a result, rows of the dataset are not independent but with dependencies clustered by user. To see this, each user's ``baseline'' measurement is captured as a random effect. The same user in different period will share the same ``baseline'' random effect, therefore resulting in dependency. Mixed effect model effectively takes advantage of this and is able to estimate the variance of the random effect while reducing the variance of average treatment effect. In the case of crossover design, the model can take advantage of the positive correlation between the two periods of the same user, which improves accuracy in the estimation of treatment effect, similar to the illustration we derived in Section~\ref{sub:crossover_design}. Treatment effect can be modeled as either a fixed effect or random effect\footnote{When there are only two measurements for a subject like crossover design, modeling treatment effect and user ``baseline'' both as random effect is unidentifiable. But the model can be fit if there are more measurements per subject.}. If our interest is the average treatment effect, we can model it as a fixed effect. Note that modeling treatment effect as fixed effect does not mean we need to assume it is fixed, which in general is not since different subjects react to the treatment differently, but rather because the focus here is the mean of the random treatment effect, not the variance of the random treatment effect. One can still fit the model with random treatment effect and the results generally agree, though fixed effect is believed to be more robust against model assumptions; see \citet{Wooldridge2012}.

We point out two issues of using traditional mixed effect model, and claim that FORME is a better alternative on axes of flexibility and scalability. 

First, linear mixed effect model (and also generalized linear mixed effect model) is a family of parametric models, and relies on full knowledge of the likelihood function to perform parameter fitting. This means the model need to rely on distributional assumptions such as normality. In particular, all random effects are typically modeled as normally distributed or jointly normally distributed. And noise $\epsilon$ need to be either i.i.d normal or the modeler needs to provide a known covariance matrix. These assumptions are indispensable in the theory and pivotal in the fitting of the model. For our application in online A/B Testing, many of these assumptions are inappropriate. To name a few, for a metric like revenue per user, it is inappropriate to model the user ``baseline'' revenue per week as normally distributed due to its large skewness. Also the noise term $\epsilon$ is hard to justify to be truly independent of other random effect. A heavier user might have bigger ``baseline'' revenue, and also bigger noise, and bigger (or smaller in some cases) treatment effect. It also assumes data are missing at random. Modelers of linear mixed effect model will need to modify the model by making random effects jointly random, or including more interaction terms. However the more complicated the model, the more questions on model assumptions will arise. We show in Section~\ref{sec:results} through simulation study, linear mixed effect model fitted in R package lme4\citep{Bates2012} could result in biased estimation of the average treatment effect when there is correlation between data missing pattern and user random effect. 

Second, fitting mixed effect model could be expensive. Available packages in SAS or R are based on fitting MLE or REML(restricted maximum likelihood). In either case, much effort is taken to estimate the variance of random effect(s) or covariance matrix if they are jointly random. Fitting algorithm takes the full dataset with each row representing a measurement. In online A/B testing, where tens of millions of users are involved, this dataset could be large. In model fitting, each iteration requires some operations on this full dataset. Making the efficiency of model fitting a concern in big data scenario. To the authors' best knowledge, there is no literature on the topic of big data implementation of linear mixed effect model. In our experience FORME is 1 to 2 magnitudes faster than lme4 with much less memory footprint even without map-reduce type parallelism.  

In the remaining of this section, we explain why FORME is both more scalable and flexible than linear mixed effect model.

\subsection{FORME is Scalable}
\label{sub:implementation_for_big_data}
Instead of modeling at the level of each individual measurements, FORME sees the problem from a higher level and take advantage of big data. Based on central limit theorem, metrics of interest in each period for treatment and control follows normal distribution. Using the same notation in Section~\ref{sec:Background}, this multivariate normal random vector is denoted by $\xbar_i, \ybar_i, i=0,\dots,2$, with mean $\vec{\beta}(\vec{\lambda})$ and certain covariance matrix. These metric values are correlated with each other via common user level random effects modeled explicitly in linear mixed effect model but not in FORME. This is because when our interest is only in the average treatment effect, the estimates of those random effects are irrelevant. Instead, FORME sees the average treatment effect $\delta$ as just one parameter in the mean vector of the metric values $\vec{\beta}(\vec{\lambda})$. That is, when modeling metric values directly using multivariate normal distribution with parameters in the mean vector, all the complexities involving the structures of the random effects are buried in the covariance matrix of multivariate normal and we are left with a simple task, which is to estimate the parameters $\vec{\lambda}$ of this multivariate normal. 

FORME estimates $\vec{\lambda}$ by fitting MLE. The use of asymptotic statistics also guarantees that the estimates are normally distributed with covariance matrix derived from Fisher's Information \citep{VanderVaart2000}. Note that the scale of this step is much smaller than the MLE fitting of a typical linear mixed effect problem. FORME only need to fit a multivariate normal with small dimension, typically smaller than 12 (6 in a crossover design: treatment and control for each of the pre-experiment, period 1 and period 2.)

The main computation burden is therefore in the estimation of covariance matrix. Fortunately, this step only involves estimation of pair-wise covariance between metric values, and they all can be map-reduced with one pass of the data. To handle missing data and general form of metrics (as a continuous function of other metrics), $delta$-method can be employed (Section~\ref{sec:missing_values}). The application of $delta$-method only involves slightly more complicated covariance matrix so we need to estimate more covariance pairs in one map-reduce pass of the data, inducing negligible increase in complexity. 

\subsection{FORME is Flexible}
\label{sub:flexible}
FORME is not only scalable but also more flexible. Because FORME doesn't explicitly model random effects as linear mixed effect models do, FORME makes no distributional assumptions on random effects and noises $\epsilon$. FORME also make zero assumption on missing data pattern. FORME needs  only one critical assumption , i.e. that central limit theorem is applicable, which is rarely violated in online A/B testing, since  traffic size is large enough even for the most highly skewed metrics such as Revenue \citep{Kohavi2014SevenRules}. Specifically, FORME can be applied to all these cases:
\begin{compactenum}
\item Data can have arbitrary missing pattern. In other words not assumptions on missing at random. 
\item Treatment Effect is random.
\item Treatment Effect and user random effect (baseline) are not independent.
\item Noises $\epsilon$ are not i.i.d.
\item Noise and random effects are not independent.
\item Interactions. (e.g. treatment and control have different user random effect distribution, etc.)
\end{compactenum}

To close this section, we make the final remark that the flexibility of FORME really comes from its simplicity, comparing to linear mixed effect model. We believe FORME is also easier for practitioners to understand. The cost of FORME to put less assumptions than mixed effect model is the expectation that when mixed effect model assumptions hold, FORME estimate could possess larger variance than mixed effect model estimate. Next we'll explore these through simulation study.

\section{Results} 
\label{sec:results}

\subsection{Simulation from Known Distributions} 
\label{sub:simulation_results}
We compare variances reported from our FORME produces to the traditional linear mixed model under various simulation assumptions. As illustration we used the crossover design. We simulate a total of $2N$ users,  where $N=10000$ and randomly split them into two treatment groups. 
\begin{equation*}
 X_{ij}=\mu + \delta_{ij} + u_i + \epsilon_{ij}, \epsilon_{ij}\sim N(0,\sigma^2) 
\end{equation*}
where $i$ is index for user and $j$ for time period. $\epsilon_{ij}$ represents  random noises and $u_{i}$ represents random user ``baseline'' effect. $\delta_{ij}$ is the treatment effect for user $i$ in period $j$ (0 if not in treatment). In this model, the between period correlation is then $\frac{\sigma^2}{\sigma_u^2+\sigma^2}$. If user $i$ is in treatment for time period $j$, $\delta_{ij} \sim N(\delta, \sigma_{\delta}^2)\times p_{i}$, where $\delta\times E(p_i)$ is the ground truth average treatment effect size, $p_{i}$ is a continuous value between 0 and 1, and it represents the user's activity level. We designed $p_{i}$ to be correlated to $u_i$. This way we allow treatment effect to vary by how frequent a user visits the site. Finally we allow $X_{ij}$ to be missing with probability of $\max(90\%, 1-p_{i})$. This is intuitive since a less active user would be missing more often. Note that in this simulation study we know exactly what the true average treatment effect is. We simulated this process $K=10000$ times so that we can have a good estimate of the ground truth variance of treatment effect estimated by FORME and mixed effect model (lme4). We want to learn the following for both FORME and lme4 from this simulation study: 1) is estimate unbiased, 2) is variance estimation correct. If both methods are unbiased, then we want to know which one has smaller variance. Without loss of generality, we used $\mu=0$, $\sigma=4$, $\sigma_u=2$ or 4, $\delta=10$, and $\sigma_{\delta} = 0.1\sqrt{12}$. We chose 5 simulation conditions as the following:
\begin{compactenum}
\item Normal noise, no treatment effect, normal user random effect $u_i \sim N(0, \sigma_u^2)$
\item Normal noise, no treatment effect, Poisson user random effect $u_i \sim Poisson(\sigma_u^2) $
\item Normal noise, with random treatment effect that is correlated with user random effect: $N(\delta, \sigma_{\delta}^2)\times p_{i}$
\item Noise is correlated with User activity level: $\sigma =2 \times p_i$
\item Noise is correlated with User activity level: $\sigma =4 \times p_i$
\end{compactenum}
and all conditions have roughly 50\% user missing in each period. 

First of all in condition 3 when there is a random treatment effect, we found lme4 consistently gave biased estimation(when ground truth effect is 6.6, FORME estimates are very close to ground truth while lme4 always gave biased estimation around 7.2). This is because lme4 relies on the assumption of missing at random, and it is violated as random effect is negatively correlated to the chance of missing. We believe this is a fundamental problem issue with mixed effect model as missing data pattern is often correlated with some underlying user characteristics that is correlated with user's response to treatment. One might argue that mixed effect model can be fixed by throwing more interaction terms. However in practice more complex models are often not identifiable (parameters more than data points) and they only makes more assumptions. We also noted that except for condition 3, lme4 provided unbiased estimates for condition 2, 4 and 5 where some assumptions in mixed effect model are violated. We believe central limit theorem also helped in this case for lme4 to stay unbiased. But bias in condition 3 seems to be more fundamental. We leave a more thorough study of the bias of lme4 with violation of different assumptions in future work. 

We also compared variance in LME and FORME under the crossover model below in Figure \ref{fig:lme_forme}. Both FORME and lme4 provided very good estimation of variance. And also as expected FORME pays a price for its flexibility and almost ``model free'' as variance from lme4 estimations are generally smaller. The variance gap is bigger when missing rate is higher and between-periods correlation is higher. Although not shown, in the conditions when either there is no missing data, or the correlation is 0, FORME and lme4 estimates have the same variance. Although lme4 estimate has smaller variance, its potential bias is a show-stopper since for treatment effect estimation a low variance estimate is not useful if biased. 

\begin{figure}[!htbp]
\begin{center}
\includegraphics[width=0.42\textwidth,height=0.25\textwidth]{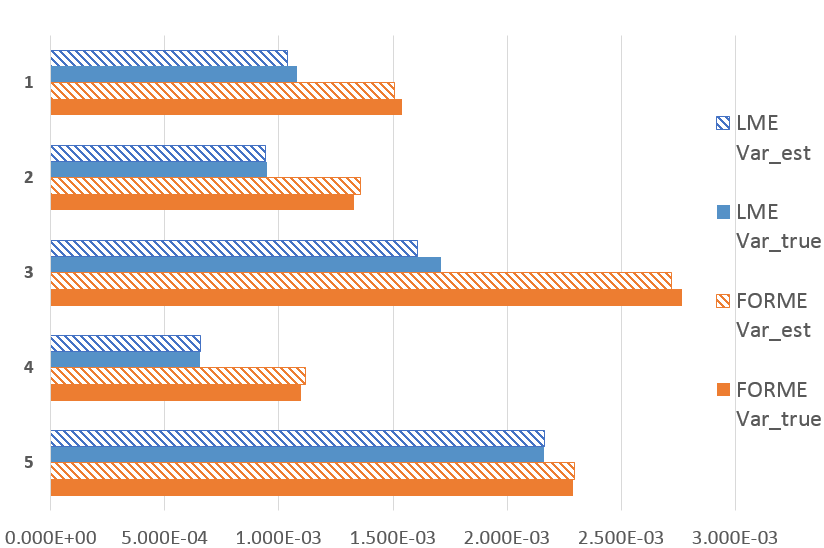}
\end{center}
\caption{Effect Variance in LME and FORME}
\label{fig:lme_forme}
\end{figure}

\subsection{Simulation from Empirical Data} 
\label{sub:empirical_results}
\begin{figure}[!htbp]
\begin{center}
\includegraphics[width=0.51\textwidth,height=0.22\textwidth]{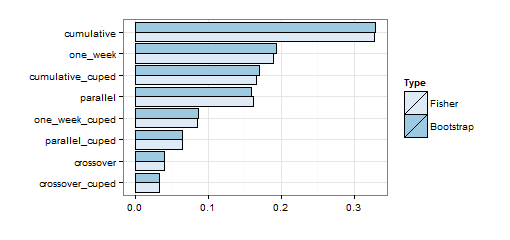}
\end{center}
\caption{Effect Variance from Fisher's Information and Bootstrap method}
\label{fig:Boot}
\end{figure}

Next we randomly sampled from our in-house data a small subset of $N=1250$ users, randomly split the users into equal sized subsets, and applied various designs. We then simulate $K=10000$ bootstrap samples (with replacement) from this dataset, fit FORME and report estimated MLEs. The variance based on these MLEs are then compare to the variance estimated from Fisher Information using the full dataset. Figure \ref{fig:Boot} shows the two agrees well. Note the cumulative effect had different effect size from the rest of the designs. For this particular metric, using CUPED results in roughly 50\% reduction in variance. Crossover design shows reduction of around 50\% compare to the parallel design. 

\subsection{Real Experiments} 
\label{sub:real_experiments}
\begin{figure}[!htbp]
\begin{center}
\includegraphics[width=0.5\textwidth,height=0.20\textwidth]{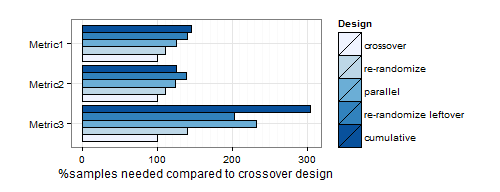}
\end{center}
\caption{Percent samples needed to achieve the same sensitivity for four metrics. Baseline is the crossover design.}
\label{fig:RealExpt}
\end{figure}

Finally, we report results from three typical metrics in one of our real experiments. Here we used percent change as the effect size. This way, weekly effect size is comparable to cumulative effect size across two weeks. The variance of effect size therefore indicates sample size needed to achieve the same sensitivity. Figure \ref{fig:RealExpt} displays the percent samples needed to achieve the same sensitivity for three metrics using various models, with the crossover design as baseline. Therefore crossover design had value of 100. All models included CUPED since pre-experiment data always exists and is free. The crossover design consistently had the fewest samples needed. Next the re-randomize design had value between crossover and the parallel design. Cumulative design follows. When the re-randomize model includes a leftover effect, the samples needed can be larger than cumulative design for metric 2. Note that compared to the previous benchmark, the cumulative design, the crossover design can save up to 2/3 the traffic for metric 3, while for others, the traffic savings is in the 30-40\% range. This is due to inherent difference in week to week correlation in different metrics. Note the drastic reduction in variance for such metrics means the same feature can be tested with only 1/3 of the original traffic!

\section{Practical Considerations} 
\label{sec:Practical_considerations}







At the design stage, we face a few choices under the same framework of repeated measures design. Experimenters should use domain knowledge and past experiments to inform the design. This is rather an art than pure science. Here we give guidelines according to our own experience. 

\subsection{Recommended Work Flow} 

\label{sub:recommended_work_flow}

Due to the flexibility in a two-stage setup in repeated measures design, we can use the information gathered in the first stage to inform procedures in the next stage. We recommend using crossover design as validation stage experiments, for which we already have gathered exploratory directional data. If the first stage already result in statistical significance in KPI, we may choose to terminate the experiment already. However in practice, we generally recommend running the experiments long enough to gain enough power for not only the KPI, but other metrics designed to monitor data quality and serve as guardrail against unexpected changes. 

Otherwise, in running the second stage, we can use domain knowledge to inform about carry over effect. If historical experiments in similar feature iterations indicate potential carry over effect, we recommend running a complete 4-group crossover experiment, so we can directly estimate carry over effect. Otherwise, we recommend using the 2-group crossover design to achieve the maximum power for KPI. If we are not sure, it is still possible to leave a few days' ``wash-out'' period after completing the first stage, and see if any carry over effect can be observed. 

\begin{compactitem}
\item \textbf{No swapping}: When it is critically important to ensure consistent users experience, such as changing the entire layout of a site, it may not be desirable to show users the new site for a week, and then swap them back to the old site. The experience may be too jarring to users and hurt the brand. In such cases, we do not recommend re-assigning treatment variants half way through. 
\item \textbf{Crossover}: Relatively small changes that are less directly noticeable are better candidates for treatment swapping. If similar experiments from the past, or earlier exploration data do not indicate the presence of carryover effect, the crossover design can be employed. 
\item \textbf{Re-randomized}: If we suspect the presence of carryover effect, the re-randomized design enables us to measure it directly and should be used here.
\item \textbf{Wash-out and decide}: If we have little information to judge carry over effect, we can run the first week of the experiment, and then leave a few days as a ``washout'' period. The next stage is data driven. Using such data we can estimate the carry over effect explicitly. 
       \begin{compactitem}
              \item If there is no significant carry over effect, proceed as the crossover design. 
              \item Otherwise, proceed as the re-randomized design. 
       \end{compactitem}
\end{compactitem}

Having collected experiment data, they can then analyzed in the following work flow to achieve the most power.
\begin{compactitem} 
\item[] \textbf{ No swapping:}       
        \begin{compactitem}
              \item Test equivalence of treatment effect across time
              \item If they are equivalence, report treatment effect in the ``per time unit" metric values by analyzing using the parallel model, including pre-experiment data. 
              \item Otherwise, analyze only cumulative effects, and including pre-experiment data. Note this is CUPED. 
       \end{compactitem}
\item[] \textbf{Crossover design:}
       \begin{compactitem}
              \item Test equivalence of treatment effect across time
              \item If they are equivalence, report treatment effect in the ``per time unit" metric values by analyzing using the crossover model, including pre-experiment data. 
              \item If, however, unexpected significant difference is found, you have several choices
                     \begin{compactitem}
                            \item Report the two treatment effects separately 
                            \item To understand the difference properly, another phrase of the experiment can be added, using re-randomized design. With a total of three weeks' data, we can see whether the treatment effect difference is due to true week-to-week to difference, and study its trend, or due to carry over effect. 
                     \end{compactitem}
       \end{compactitem}
\item[] \textbf{ Re-randomized design:}
       \begin{compactitem}
              \item Test equivalence of treatment effect across time and presence of carryover effect
              \item Reduce the model if any of the effects are not statistically significant, and report treatment effect.
       \end{compactitem}
\end{compactitem}

This carries the subtle difference of reporting a treatment effect in the entire duration of the experiment, versus that per time unit (a week here). We argue that as long as weekly treatment effects are stable over time, reporting weekly effect is intuitive, easy to understand, and easy to compare across different experiments. In real life, various things can happen during an experiment, and we may end up with an experiment that ran only in partial weeks. In these cases, reporting treatment effect in the entire duration will be better than throwing away data or ignore weekdays difference. 



\subsection{Sample Size Considerations} 
\label{sub:sample_size_considerations}
While direct estimation in sample size is difficult in the linear mixed model, in practice there is an easy work around. In the traditional design, using CLT, with a simple two-sided test for $H_0: \delta=0$, sample sizes can be easily calculated. 
\begin{equation*}
       n = (z_{1-\alpha/2}+z_{1-\beta})^2/\frac{\delta^2}{Var(\delta)}
\end{equation*}
where $\alpha$ is the allowed false positive rate, usually 0.05, and $1-\beta$ is the desired power, usually at $80\%$ to $90\%$. 

From historical data we can record the amount of variance reduced for each metric. The magnitude is determined by inherent variance in the metric, and correlation across time periods, both of which are observed to be fairly stable across many experiments. Suppose the variance for metric X in crossover experiment is $k\%$ that in the conventional t-test. If $N$ subjects are required to detect a change of $\delta\%$ in t-test with, say, $80\%$ power, then $k\%N$ is the reduced sample size to achieve the same power. 

\section{Discussions and Future Work}

\subsection{Extending to more frequent swaps} 
\label{sub:extending_to_more_frequent_swaps}
The crossover design achieves sensitivity by exposing users to both treatment variants in sequence, by swapping the treatment assignment once during the experiment. Using each subject as his or her own control and this design to account for within-subject variance. A natural extension of the idea is to swap treatment groups more than once. Essentially, this changes to a more granular randomization unit, from users to page views. Exploratory work shows this indeed achieves further variance reduction. 

However, this also raises the concern for inconsistent user experience, diminished treatment effect size, stronger learning effect, and lack of a longer term measure. Despite these concerns, it remains a valuable option in early stage experiments to quickly select promising features for further iteration. 


\subsection{Limitations and concerns} 
\label{sub:limitations_and_concerns}
Due to user behavior differences between weekday and weekends, we usually recommend running each phase of the cross-over design for at least a full week. A crossover experiment then requires two complete weeks to gather data, which hinders agility. Another limitation is that for very highly visible features like changing prominent UI features, such swapping may not be desirable since it may confuse the users. Finally, not all features can be tested this way, as there might be a ``learning'' effect, where we can't have the users exposed to treatment ``unlearn'' the feature, while having controls naive to the treatment. For example, if the website provides new features and personalized content to signed in users to encourage higher rate of signing in and staying signed in. These users cannot then be forced to logout into the control group. \citet{Ma2011} shows one interesting case where crossover design can be extended to tackle this issue.

\subsection{Further Improvements of FORME}
We've shown in Section~\ref{sub:simulation_results} that mixed effect model via lme4 provides a competing estimate of the average treatment effect that could be biased when missing data pattern correlate with user random effect, but often with smaller variance than FORME. We noted that FORME has to pay some price to be more flexible and robust, similar to nonparametric model usually is less efficient than their parametric counterparts. However we suspect that efficiency of FORME can be further improved to match the efficiency of mixed effect model even under perfect mixed effect model assumption. Such improvement would be very desirable. But even without such improvement we believe the bias when there is missing data that is not missing at random is a big issue for mixed effect model to be adopted in online controlled experiment. And FORME should be used instead.

\bibliography{library}

\nocite{Ma2011,VanderVaart2000,Kohavi2014SevenRules,puzzlingOutcomes,deng2013cuped,DengTwoStage,bakshystatistics,Bates2012,Bates2012a,romano2005testing,Xu2009}

\end{document}